\documentclass[onecolumn,showpacs,floats,floatfix,aps,prl,groupedaddress]{revtex4}
\usepackage{graphicx}
\usepackage{url}
\usepackage{amsmath,amssymb,amsfonts}
\usepackage{subfigure}

\begin{document}

\title{\textbf{Directed Motion of Elongated Active Polymers}}

\author{Mew-Bing Wan$^{1}$, YongSeok Jho$^{1,2,*}$}

\affiliation{{$^1$}Asia-Pacific Center for Theoretical Physics, Pohang, Gyeongbuk 790-784, South Korea\\
{$^2$}Department of Physics, POSTECH, Pohang 790-784, South Korea\\
{$*$}ysjho@apctp.org}

\date{\today}

\begin{abstract}
Previous work shows that a net directed motion arises from a system of individual particles undergoing run-and-tumble dynamics in the presence of an array of asymmetric barriers. Here, we show that when the individual particle is replaced by a chain of particles linked to each other by spring forces (polymer), the rectification is enhanced. 
It is found that the rectification increases when the number of particles in each polymer, as well as its length, increases. In addition, the rectification increases when 
the size of the opening between neighboring funnel tips, $l_o$, decreases. Interestingly, if the conformal entropic difference exceeds the thermal diffusion, net directed motion is observed even when the run-and-tumble dynamics approaches Brownian motion. 
Also, when the inelastic collisions between the particles and the barriers are replaced by elastic collisions, a \textit{reversed} rectification is observed.	


\end{abstract}

\maketitle

\section{I. Introduction}

The interplay between physics and biology has recently not only illuminated important underlying biological mechanisms, but also given birth to new avenues of understanding physics, which in turn has lead to interesting applications. 
A good example concerns the recent series of works that explore the dynamics of certain species of bacteria in relation to non-Brownian diffusive motion, which can be harnessed to perform work in spite of the limitations of the 2nd Law of Thermodynamics.
It has been known for quite some time that the motion of bacteria such as the \textit{Eschericia Coli}, can be described as a random walk. 
Each bacterium self-propels in a straight line with an almost constant speed, called a "run", after which it "tumbles" by "unbundling" its flagellar motor and propels again in a straight line along a random direction \cite{Berg03}.
It is found recently that a bath of such bacteria, without imposing any external forces, exhibits a type of diffusive behavior, called rectification, when placed in an asymmetric ratchet system \cite{Galajda07}. 
This constitutes an intriguing phenomenon of diffusion even in the absence of detailed balance.
In this work, the bacteria are seen to interact inelastically with the funnel ratchets. If they collide with the funnel walls at oblique angles, they would slide along the walls until reaching either ends of the walls or until their next "tumble" events.
Subsequently, a series of works explore bacteria bath systems interacting with asymmetric saw-toothed rotors, resulting in similar rectification phenomena \cite{Angelani09,Leonardo10,Sokolov10}.

Theoretically, this behavior is first modeled in a two-dimensional system \cite{Wan08,Cates09}. In these works, the bacteria are immersed in a two-dimensional box, in which an asymmetric funnel ratchet system is placed.
The bacteria are modeled as particles engaging in the previously mentioned "run-and-tumble" dynamics.
Without imposing any external force, the bacteria are rectified to the upside of the funnels after a certain amount of time. 
By the Navier-Stokes equation, such "run-and-tumble" dynamics should be describable as a time-symmetric process due to the low Reynolds number of swimming bacteria \cite{Galajda08}.
However, the randomness of the "bundling" and "unbundling" of the bacteria flagellar motors breaks time-reversal symmetry.
Therefore, a hydrodynamics approach is inadequate to give a full picture of this type of dynamics.
When the length of the "run" is reduced, rectification is reduced \cite{Wan08}, even though the sliding behavior is maintained. 
This corresponds to the experimental finding by the authors of Ref. \cite{Galajda07} that non-swimming or dead bacteria do not show any rectification. 
It can be seen that the ratio of the funnel opening over the run length has an effect on the rectification magnitude. 
When this ratio is much bigger than unity, the rectification vanishes.
This should be expected as non-swimming or dead bacteria are only subject to thermal fluctuations, in which case, thermal equilibrium and detailed balance are restored. 
On the other hand, it is found that when the sliding behavior is replaced by elastic collisions \cite{Cates09,Reichhardt11}, or by scattering, i.e., when the particles reflect off the funnel walls \cite{Reichhardt11}, the rectification also vanishes. 
It can thus be deduced that the long "run" lengths of the random walk, coupled with the time-reversal asymmetric interaction of the particles with the funnel walls, breaks detailed balance, inducing the rectification.

The fact that such matter is in a strongly non-equilibrium state is crucial to its ability to be manipulated by means of asymmetric geometries alone in generating directed motion \cite{Leonardo10}.
The novelty of this phenomenon stems from its contrast with previous observations where external forces are needed to control the random motions of the bacteria. 
In these previous cases, external forces in the form of electric or magnetic fields \cite{Speer10} or optical fields \cite{Xiao10} are applied in order to generate the directed motion.
The phenomenon of rectification with an asymmetric geometric ratchet system therefore opens up the possibility of building self-propelled machines just via the intrinsic dynamics of a bacterial bath. 
In addition, the ordered behavior that emerges from the dynamics of the bacterial bath can be qualitatively and quantitatively controlled 
by varying the geometry of the ratchet system \cite{Galajda08,Wan08,Cates09} or just by controlling the amount of oxygen fed to the bacteria \cite{Sokolov10}.

So far, the theoretical modeling have only focused on point particles for the sake of simplicity, and the fact that they can reproduce the basic characteristics of the rectification phenomena seen in experiments.
However, since bacteria are non-point-like, the rectification phenomena may exhibit richer characteristics in the presence of the asymmetric ratchet systems. 
In addition, since different bacteria species or cells have non-negligible differences in their aspect ratios, it is possible that each of these different species or cells rectifies differently.
In this paper, we show that when the individual particles are replaced with elongated polymers consisting of particles linked to each other by spring forces, the rectification exhibits appreciable quantitative and qualitative differences compared to the individual particle/single-monomer case. 
We invoke a flux balance argument for the rectification of the single-monomer case and an entropic argument for the quantitative differences that arise in the polymer case in comparison with the single-monomer case.  

\section{II. Setup and Results}

\begin{figure*}
\subfigure[]{
\includegraphics[scale=0.5]{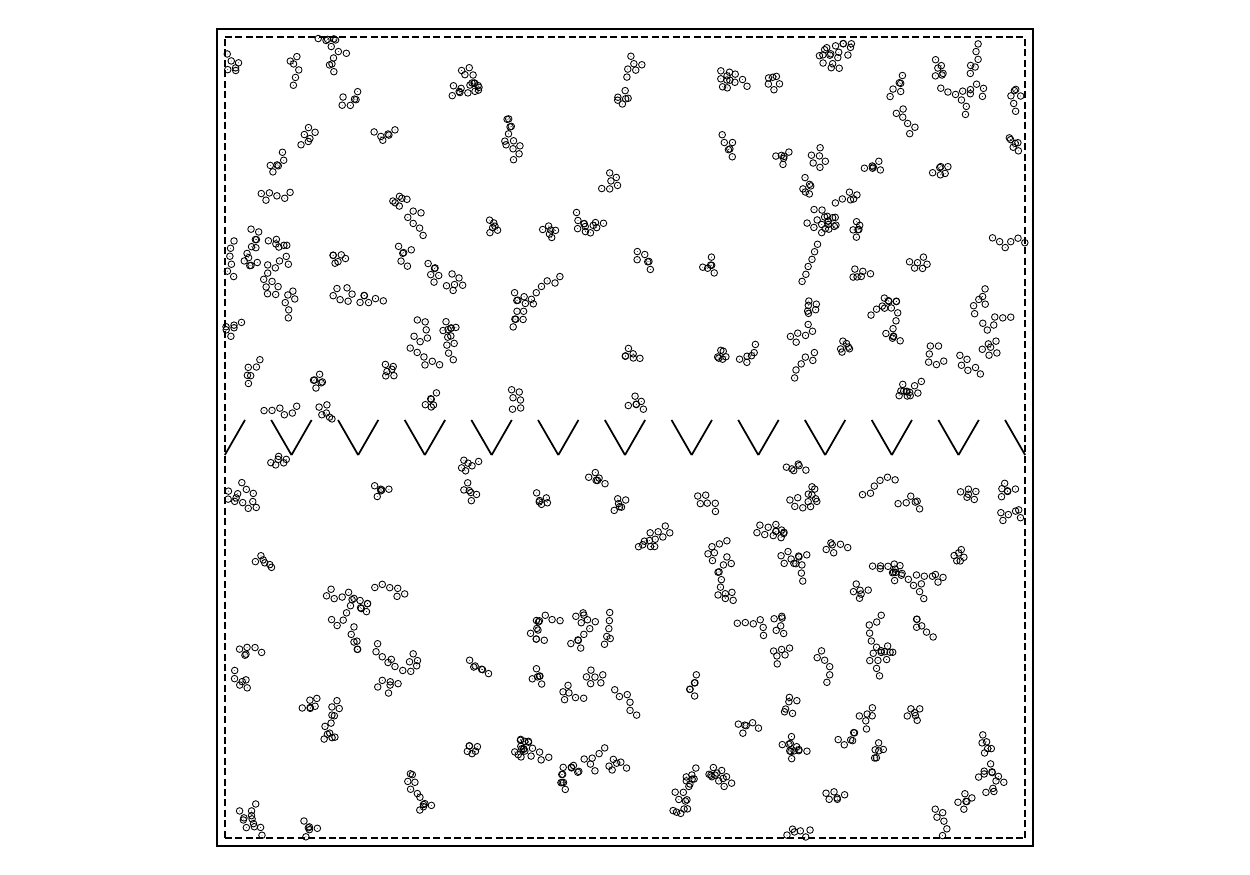}
}
\subfigure[]{
\includegraphics[scale=0.5]{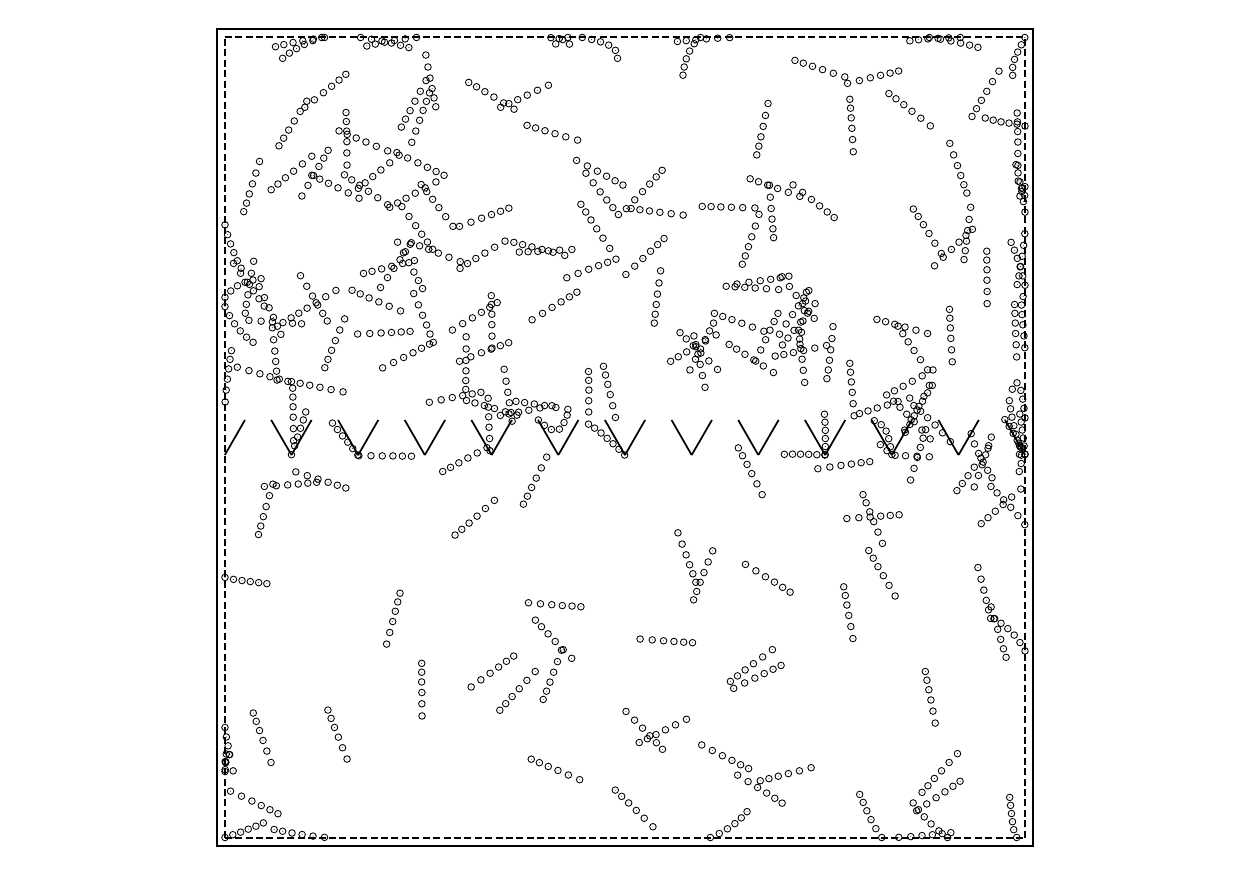}
}
\caption{Simulation box for the six-monomer polymer case at time, (a): $t=0$, and (b): $t=6\times 10^6$.}
\label{fig:2box}
\end{figure*}

Our system consists of a two-dimensional box of dimensions $x\times y=L\times L$ with an array of funnel-shaped barriers placed along the $x=L/2$ line separating the box into chambers $1$ and $2$. 
The array of funnel-shaped barriers consists of $N_B$ barriers of length $L_B$ arranged in $V$ shapes with opening angles of $2\theta$ and distance $l_d=2L/N_B$ from each other.    
The box is populated with $N_{poly}$ number of polymers with $N_{mon}$ number of monomers in each polymer. 

Initially, $1200$ particles are distributed evenly in chambers $1$ and $2$. The dimensions of the box are fixed with $L=99$ in which we set up a system of $28$ barriers of length $5$ with angles tilting from the vertical axis set uniformly at $\theta=30$ degrees (Fig.~\ref{fig:2box}).      
Following Ref.~\cite{Wan08}, the dynamics of the particles is given as an overdamped Langevin equation as follows:
\begin{equation}
\label{eq:EOM}
\eta\frac{d{\bf R}_l}{dt}={\bf F}_{Bl}+{\bf F}_k({\bf R}_{lj})+{\bf F}_{ml}(t)+{\bf F}_{Tl},
\end{equation}
where $\eta=1$ is the phenomenological damping term. No hydrodynamics is considered in our model. 

${\bf F}_{Bl}$ describes the force that repels the particles or polymers in the direction perpendicular to the barrier, as well as in radial directions from the ends of the barriers:
\begin{equation} 
\label{eq:FB} 
{\bf F}_{Bl} = \sum\limits_{i=0}^{N_B}\Bigg[\frac{F_B r_1}{r_B}\delta(r_1){\bf\hat{R}}^{\pm}_{li}+\frac{F_B r_2}{r_B}\delta(r_2){\bf\hat{R}}^{\bot}_{li}\Bigg],
\end{equation} 
where $F_B=30$, $r_B=0.05$, $r_1=r_B-R^{\pm}_{li}$, $R^{\pm}_{li}=|{\bf R}_{l}-{\bf R}_{Bi}\pm L_B{\bf\hat{p}}_{\parallel i}|$, ${\bf\hat{R}^{\pm}}_{li}=|{\bf R}_{l}-{\bf R}_{Bi}\pm (L_B/2) {\bf\hat{p}}_{\parallel i}|/R^{\pm}_{li}$,
$r_2=r_B-R^{\perp}_{li}$, $R^{\perp}_{li}=|({\bf R}_{l}-{\bf R}_{Bi})\cdot{\bf\hat{p}}_{\perp i}|$, ${\bf\hat{R}^{\perp}}_{li}=|({\bf R}_{l}-{\bf R}_{Bi})\cdot{\bf\hat{p}}_{\perp i}|{\bf\hat{p}}_{\perp i}/R^{\perp}_{li}$.
Here, ${\bf R}_{l}$ is the position of the particle, whereas ${\bf R}_{Bi}$ is the position of the center of the barrier. 
${\bf\hat{p}}_{\perp i}$ and ${\bf\hat{p}}_{\parallel i}$ are respectively the unit vectors perpendicular and parallel to the barrier. 
The first term turns on when the radial distance of the particle from the tip of the barrier is less than $r_B$.
Similarly, the second term in Eq.~\eqref{eq:FB} turns on when the perpendicular distance between the particle and the axis of the barrier is less than $r_B$.
The first term therefore gives the repelling force in radial directions from the barrier tips. The magnitude of this force is proportional to the distance that the particle traverses past the radius at the barrier tips, i.e., $r_1$.
The second term gives the repelling force in the direction perpendicular to the barrier axis, with a magnitude proportional to the distance the particle traverses past the half-thickness of the barrier, i.e., $r_2$.
With this setup, the barrier force is able to cancel the component of the particle's ballistic force perpendicular to the barrier upon the particle's approaching the barrier. 
Therefore, the particle slides along the barrier with a partial alignment.
In addition, this barrier force is able to cancel the component of the particle's ballistic force that is in a radial direction from the ends of the barrier.
This is so that two barriers will be smoothly connected at the bottom tip of the funnel. Particles are also able to slide smoothly around the upper tips of the funnels.  

${\bf F}_k({\bf R}_{lj})$ describes the force between the monomers in each polymer, and is modeled as a spring force as follows:
\begin{equation}
\label{eq:Fk}
{\bf F}_k({\bf R}_{lj}) = k(R-l){\bf\hat{R}}_{lj},
\end{equation}
where $k$ is the spring constant, $R=|{\bf R}_l-{\bf R}_j|$ is the relative distance between the monomers, ${\bf\hat{R}}_{lj}=({\bf R}_l-{\bf R}_j)/|{\bf R}_l-{\bf R}_j|$ is the unit displacement vector between the monomers, and $l$ is the equilibrium distance between the monomers,
which is set at $1$.
The first monomer of each polymer is driven by a ballistic force of magnitude $|{\bf F}_m|$ that induces a run length of $l_r=ndt|{\bf F}_m|$, every $n$ number of time steps of size $dt$ before it reorients.
The direction of the reorientation is random and mimics the rotational diffusion in this system.
The whole polymer is driven by this ballistic force, with the help of the corresponding spring forces that push and pull between the monomers.
In our model, the polymers are not only made to slide along the barriers but also to bend around when encountering the barriers.
We adjust the spring constant of the polymers so that they bend naturally around the tips of the funnels.

For an $n$ number of time steps with a step size $dt$, between tumbles, a ballistic force of $|{\bf F}_{m1}|$ induces a run length of $ndt|{\bf F}_{m1}|$ in the single-monomer case.
As the number of monomers is increased, spring forces pull in the direction opposite to that of the ballistic force of the polymer.
The total magnitude of the spring forces corresponds to the number of monomers in the polymer.
In order to induce the same effective run length as that in the single-monomer case, the ballistic force has to be increased proportionately with the number of monomers in the polymer.
Therefore, we set the ballistic force at $|{\bf F}_m|=|{\bf F}_{m1}|N_{mon}$.

We estimate the time for the last monomer to reorient in the direction of the ballistic force of the first monomer, as $\Delta t=(N_{mon}-1)l/|{\bf F}_m|$. 
For the polymers we consider below, $\Delta t$ is always less than $ndt$, the time period of each run. 
Also, the rotational diffusion of the polymer, which is characterized by the Rouse time, is shorter than $ndt$. 
As an illustration, we take the longest polymer in our study, that is the six-monomer polymer, with an effective run length of $10$, resulting from $10000$ time steps of $dt=0.0005$. In order to induce this effective run length for the entire polymer, a ballistic force of $12$ is imposed on the first monomer of the polymer. We estimate the Rouse time for this polymer to be $\tau\sim{l N_{mon}^2}/D=0.05$. We note that the Rouse time is thus smaller than the time period of each run, $ndt=5$. 
The correlation between consecutive runs of the polymers can thus be considered negligible.

\begin{figure*}
\begin{center}
\includegraphics[scale=0.85]{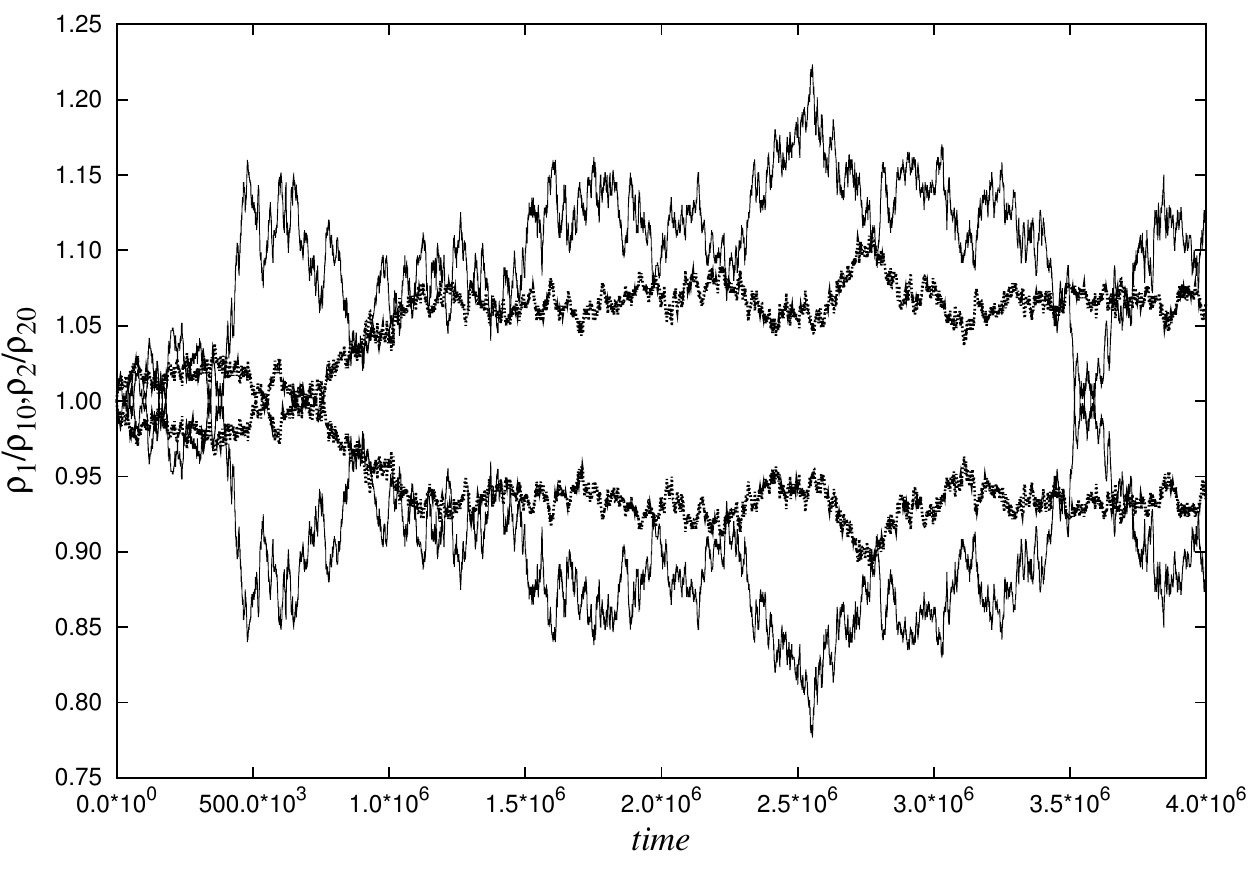}
\caption{Evolution of normalized particle densities, $\rho_1/\rho_{10}$ for chamber $1$ and $\rho_2/\rho_{20}$ for chamber $2$, comparing between the single-monomer case and the six-monomer polymer case for the effective $l_r=2$. $\rho_{10}$ and $\rho_{20}$ are the initial densities in chambers $1$ and $2$ respectively.
The dotted lines represent the single-monomer case whereas the solid lines represent the six-monomer polymer case.}
\label{fig:recm}
\end{center}
\end{figure*}

As our first set of simulations, we set the temperature term, ${\bf F}_{Tl}$ to zero. Each single-monomer or polymer is made to move $2000$ time steps of $dt=0.0005$ before randomly reorienting.
With these parameters, an effective run length of $2$, which is short enough to mimic Brownian motion, is induced for both the single-monomer and the polymer case. 
To quantify the rectification, we use the ratio $r=\rho_1/\rho_2$ ($\rho_1$ is the density of particles in chamber $1$ and $\rho_2$ is that in chamber $2$). The initial value of $r$ is $1$. 
The single-monomer case gives a rectification of magnitude $1.083$.  For the six-monomer polymer case with $200$ polymers, the rectification is increased to $1.246$.
In Fig.~\ref{fig:recm}, we show this enhancement of the rectification via the time evolution of the densities of particles in the upper and lower chambers normalized with the densities at time, $t=0$.   

\begin{figure*}
\begin{center}
\includegraphics[scale=0.85]{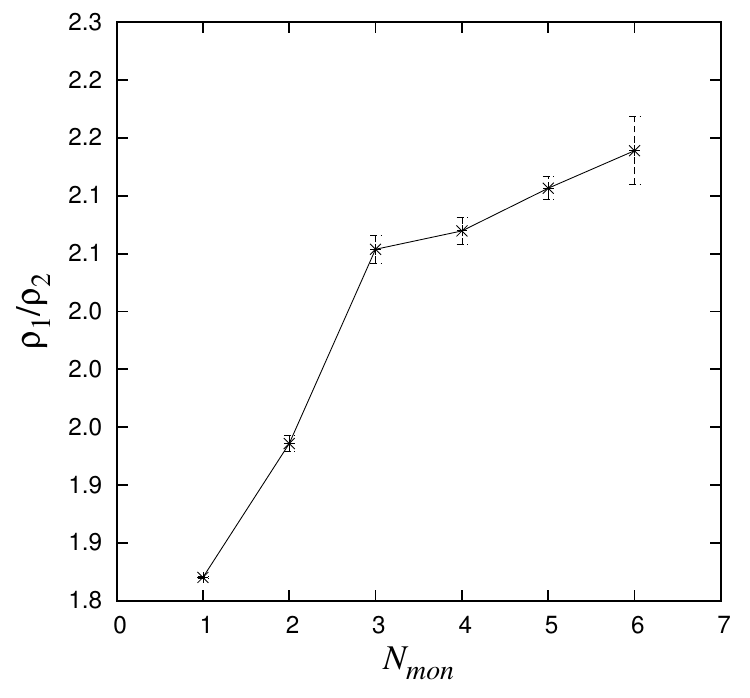}
\caption{Rectification magnitude for systems with varying numbers of monomers, $N_{mon}$, for the effective $l_r=10$, with all other parameters fixed. The error bars are obtained from the mean squared error of the rectification magnitudes of a sample of $10$ simulations.}
\label{fig:rpoly}
\end{center}
\end{figure*}

Next, we set the run length at $10$, by increasing the number of steps taken during the run to $10000$, using the same ballistic force as above. 
We then observe the change in the magnitude of the rectification by varying the number of polymers and their corresponding number of monomers in each polymer according to $N_{poly}\times N_{mon}=1200$ (Fig.~\ref{fig:rpoly}).
We find that the rectification increases almost monotonically when the number of monomers is increased starting from the single-monomer case.

\begin{figure*}
\begin{center}
\includegraphics[scale=0.85]{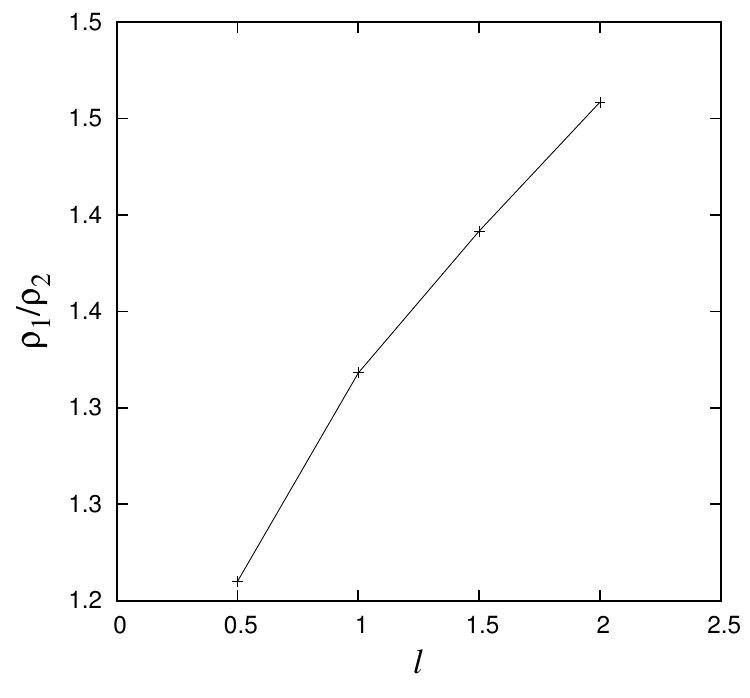}
\caption{Rectification magnitude for the six-monomer polymer system with varying equilibrium distances between monomers, $l$, for the effective $l_r=2$, with all other parameters fixed.}
\label{fig:rbl}
\end{center}
\end{figure*}

Focusing on the six-monomer polymer case with run length $2$, we vary the equilibrium distance between the monomers, $l$, and observe the effect on the rectification (Fig.~\ref{fig:rbl}). We find that as the total equilibrium length of the polymers, $l_p = (N_{mon}-1)l$ ($N_{mon} = 6$ in this case) increases, the magnitude of the rectification increases as well.  
This is in line with the above observation when the number of monomers in each polymer is increased. 

\begin{figure*}
\begin{center}
\includegraphics[scale=0.85]{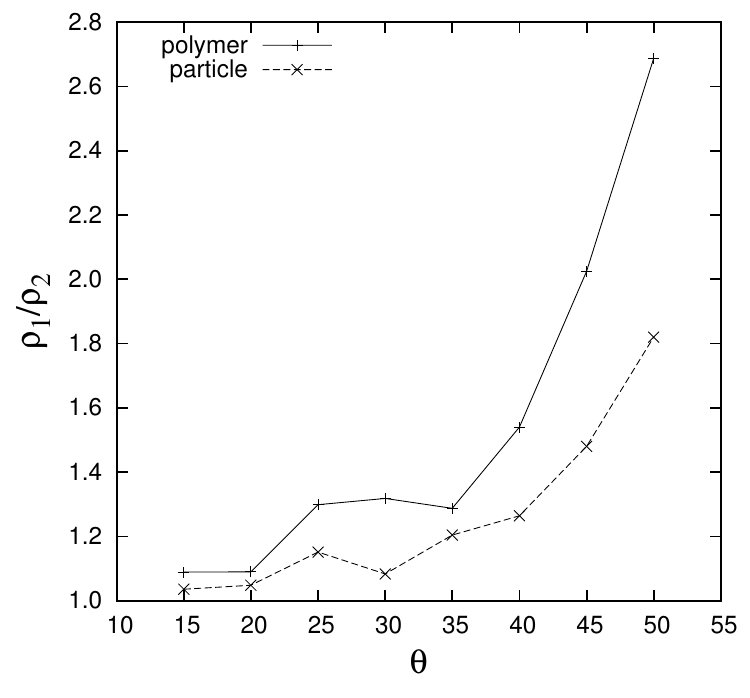}
\caption{Rectification magnitude for the six-monomer polymer system with varying funnel opening angles, $\theta$, for the effective $l_r=2$, with all other parameters fixed.}
\label{fig:rang}
\end{center}
\end{figure*}

Finally, we vary the opening angle $2\theta$ of the barriers for the same six-monomer polymer case with run length $2$ and $l=1$ (Fig.~\ref{fig:rang}). Similar to the single-monomer case, we find that as $\theta$ increases, or when the size of the opening between the barriers, $l_o=l_d-2L_B\sin\theta$ decreases, the magnitude of rectification is increased. However, the rectification for the polymer case is enhanced up to $1.4$ times compared to the single-monomer case when the tilting angle of the barriers from the vertical is increased to $55$ degrees.   

\section{III. Analyses and Discussions}

\subsection{Flux balance argument}

\begin{figure*}
\begin{center}
\includegraphics[scale=0.3]{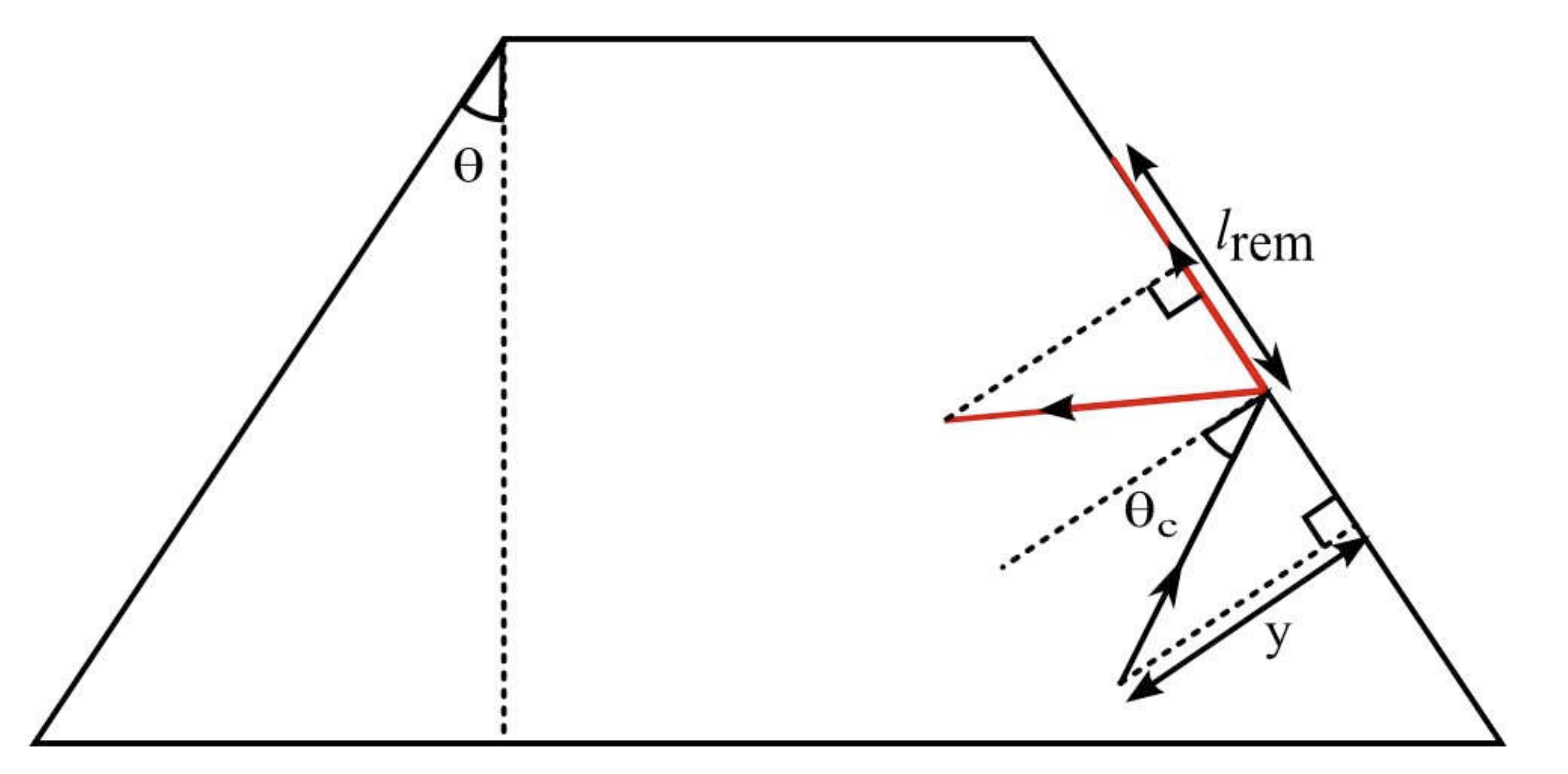}
\caption{Geometry of the space between two funnels depicting two possible trajectories of a particle colliding with one of the barriers. One trajectory is of an elastic collision and the other is of an inelastic collision, where the particle moves along the barrier after collision.}
\label{fig:delr}
\end{center}
\end{figure*}

\begin{figure*}
\begin{center}
\includegraphics[scale=0.85]{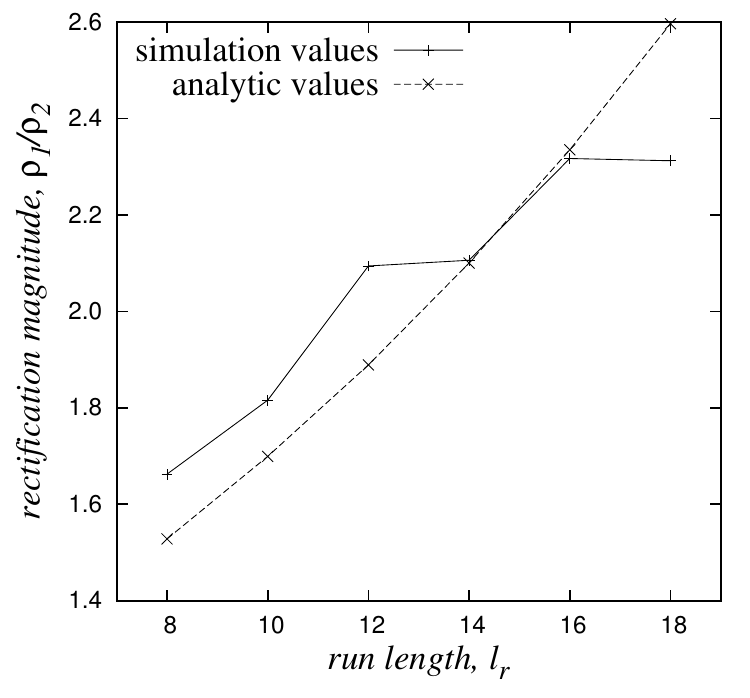}
\caption{Comparison between values of the rectification magnitude obtained from simulation and analytic analysis for the monomer case with varying run lengths.}
\label{fig:comp}
\end{center}
\end{figure*}

\begin{figure*}
\begin{center}
\includegraphics[scale=0.85]{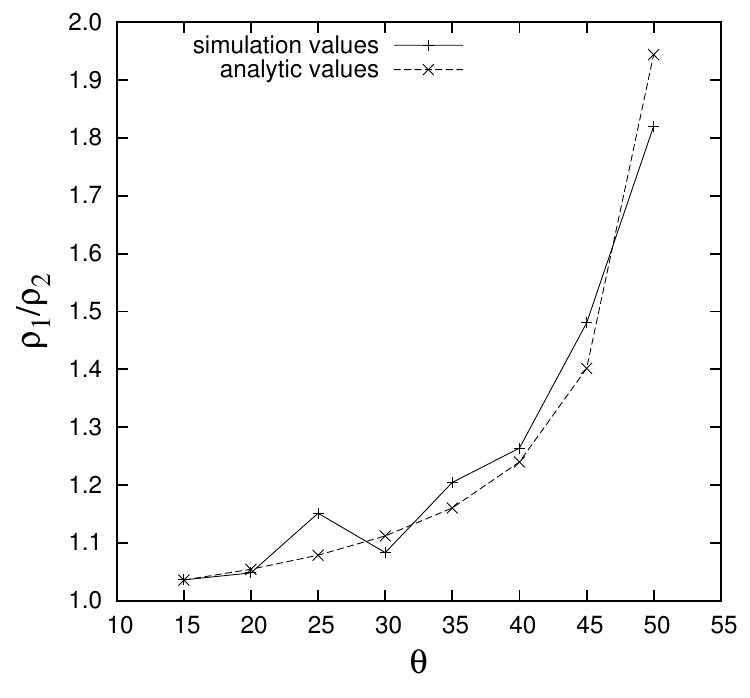}
\caption{Comparison between values of the rectification magnitude obtained from simulation and analytic analysis for the monomer case with varying barrier tilting angles.}
\label{fig:compb}
\end{center}
\end{figure*}

In this subsection, we propose an explanation of the rectification seen in the previous section using a flux balance argument. 
In the region $0\leq x<L/2-h$ and $L/2+h<x\leq L$, where $h = (L_B\cos\theta)/2$, since particles do not undergo collision with the barriers, the conservation of the particle flux, $J$, along the direction of $x$, can be written following Eq. $(5)$ in Ref. \cite{Rivero89} as:
\begin{equation}
\label{eq:convf1}
\partial_t J = -Jr_{tot} - \frac{l_o}{l_d}v^2\nabla\rho(x),
\end{equation}
where $r_{tot} = r_+ + r_-$ is the sum per unit time of the probability of particles moving from chamber 1 into chamber 2, $r_+$, and vice versa, $r_-$, $v = |{\bf F}_m|$ which can also be interpreted as the particle velocity, and $\rho(x)$ is the averaged particle number density over the $y$ direction defined as $\rho(x) = \frac{1}{\pi L}\int^L_0\rho(x,y)dy$.
The factor $l_o/l_d$ before the second term ensures that the flux change in the $x$ direction is restricted by the size of the opening between the tips of the neighboring funnels, $l_o$. 
In the region $L/2-h\leq x\leq L/2+h$, particles collide with the barriers in an inelastic way, generating a biased impulse in one direction. The gain in flux due to the inelastic collision with the barriers is described as the difference in the trajectory resulting from an inelastic collision and that from an elastic collision.  
If we define this difference as $\Delta r = r_+ - r_-$, the flux balance equation above is modified as (\cite{Rivero89}):
\begin{equation}
\label{eq:convf2}
\partial_t J = -Jr_{tot} - \frac{l_o}{l_d}v^2\nabla\rho(x) +  \frac{l_d-l_o}{l_d}v\rho(x)r_{bias},
\end{equation}
where 
\begin{equation}
\label{eq:convf3}
r_{bias} = \frac{\Delta r\cos\theta}{2hndt}. 
\end{equation}
The factor $(l_d-l_o)/l_d$ here indicates that the biased flux change in the $x$ direction is caused by the interaction of the particles with the funnels.
We calculate $\Delta r$ when the run length, $l_r$, is on a similar scale as the distance between the $V$-shaped funnels, $l_d$, as follows: 
\begin{equation}
\label{eq:convf4}
\Delta r = \frac{2 L_B\sin\theta}{l_d}\frac{1}{2\pi}\int^{l_r}_0 dy\int^{\theta_0}_0 (l_r-\frac{y}{\cos\theta_c})(1-\sin\theta_c)d\theta_c,
\end{equation}
where $\theta_0 = \cos^{-1}(y/l_r)$ and $\theta_c$ is the angle between the barrier and the trajectory of the particle heading toward the barrier.
Here, the term $l_{rem} = l_r-\frac{y}{\cos\theta_c}$ is the remainder of the trajectory of the particle colliding with the barrier, whereas the term $l_{rem}(1 - \sin\theta_c)$ is the difference in the trajectory resulting from an inelastic collision and that from an elastic collision (Fig.~\ref{fig:delr}). 

In a steady state, $\partial_t J = J = 0$. Therefore, Eq.~\eqref{eq:convf2} becomes:
\begin{equation}
\nabla\rho(x) = \frac{(l_d-l_o)}{l_o}\frac{\rho(x)}{v}r_{bias},
\end{equation}
which gives the solution for the averaged particle number density as follows:
\begin{equation}
\rho(x) = A\exp[\frac{(l_d-l_o)}{l_o}\frac{x}{v}r_{bias}],
\end{equation}
with $A$ being an integration constant.
Designating $\rho_1 = \exp[(L/2+h)/v]$ and $\rho_2 = \exp[(L/2-h)/v]$, the rectification can then be obtained as:
\begin{equation}
\label{eq:ratio}
\rho_1/\rho_2 = \exp[\frac{(l_d-l_o)}{l_o}\frac{2h}{v}r_{bias}].
\end{equation}

We next compare the rectification magnitudes obtained from this analytic analysis with those obtained from the numerical simulations for the monomer case with varying run lengths as well as barrier tilting angles.  
Fig.~\ref{fig:comp} shows that, for the case with varying run lengths, similar with the simulation results, the rectification magnitude increases almost linearly as the run length increases. 
The integral in Eq.~\eqref{eq:convf4} is found to scale linearly with respect to $l_r$.
Therefore, using Eq.~\eqref{eq:ratio}, the rectification magnitude scales like $\sim\exp(l_r)$. 
For the case with varying barrier tilting angles, Fig.~\ref{fig:compb} shows that the rectification magnitude at first increases slowly as the angle increases.
However, when the angle approaches that where the funnels block out completely the monomers on one side of the box
from the other side, the increase in the rectification magnitude is enhanced to the point of approaching an asymptote. 
The asymptote is determined by this angle of total blockage of the monomers by the funnel system, $\theta_d=\sin^{-1}(l_d/(2L_B\sin\theta)$. 
According to Eq.~\eqref{eq:ratio}, taking into account  Eq.~\eqref{eq:convf3} and Eq.~\eqref{eq:convf4}, the rectification magnitude scales as $\sim\exp\theta$ when $\theta$ is small.
When $(\theta_d-\theta)$ is small, the rectification magnitude scales as $\sim\exp(1/\sin(\theta_d-\theta))$. 

\subsection{Entropic argument}

\begin{figure*}
\begin{center}
\includegraphics[scale=0.85]{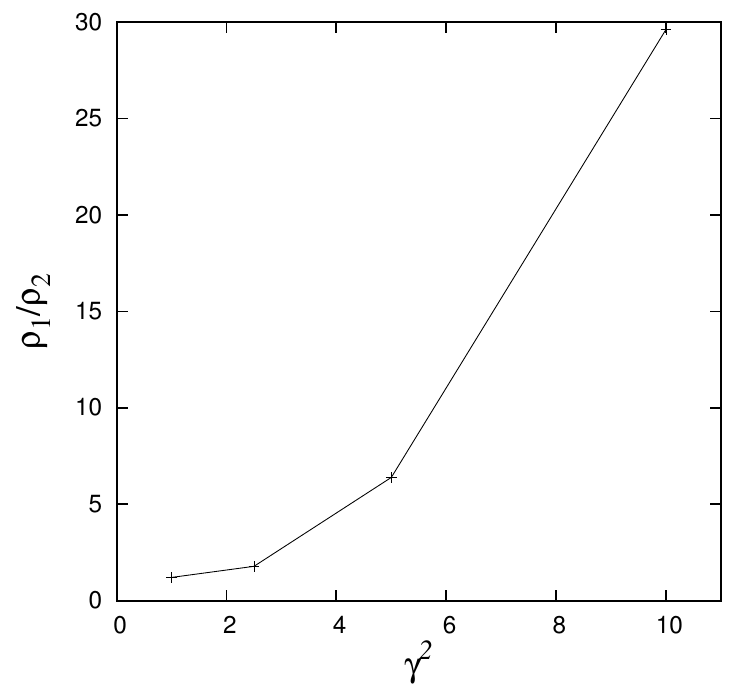}
\caption{Rectification magnitude for six-monomer polymer systems subjected only to the thermal force with varying $\gamma^2$ values.}
\label{fig:rtherm}
\end{center}
\end{figure*}

\begin{figure*}
\subfigure[]{
\includegraphics[scale=0.5]{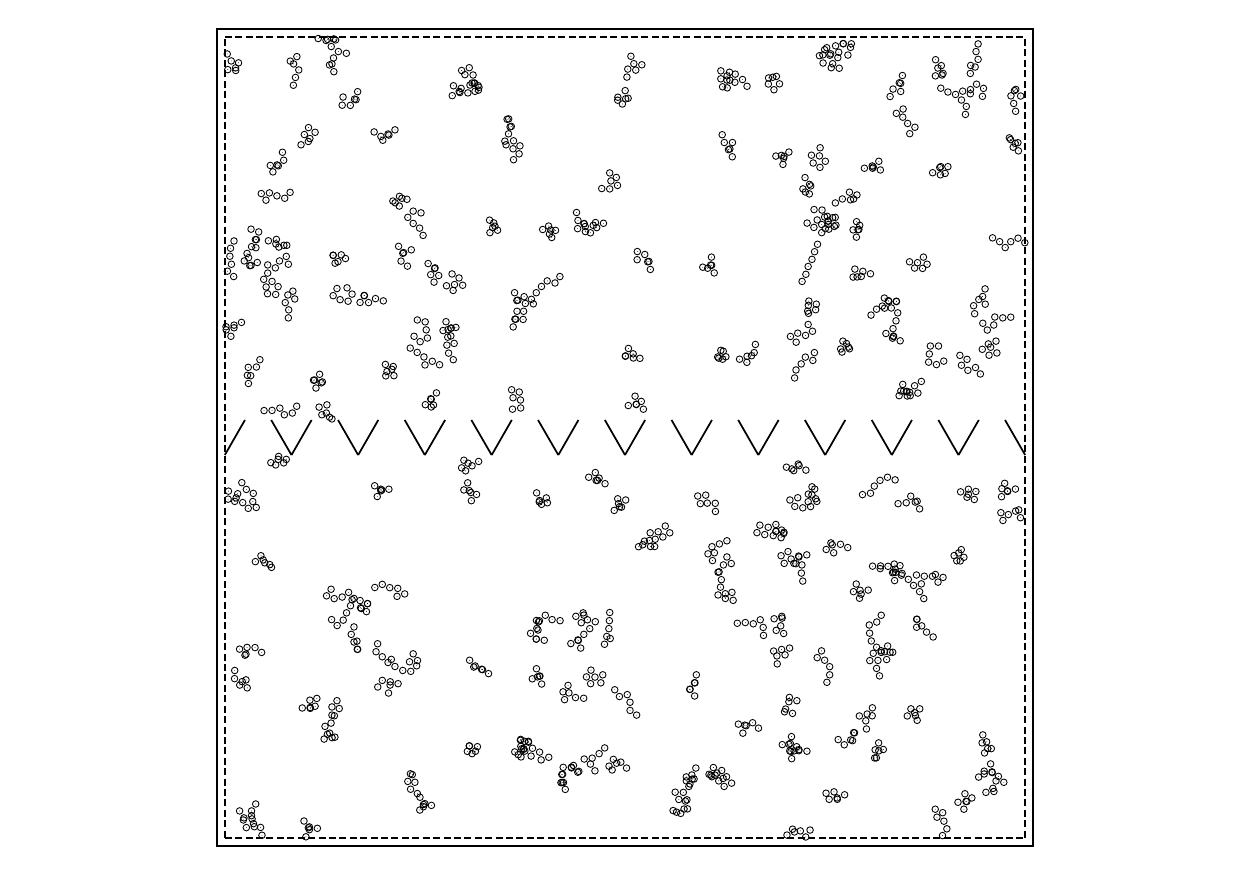}
}
\subfigure[]{
\includegraphics[scale=0.5]{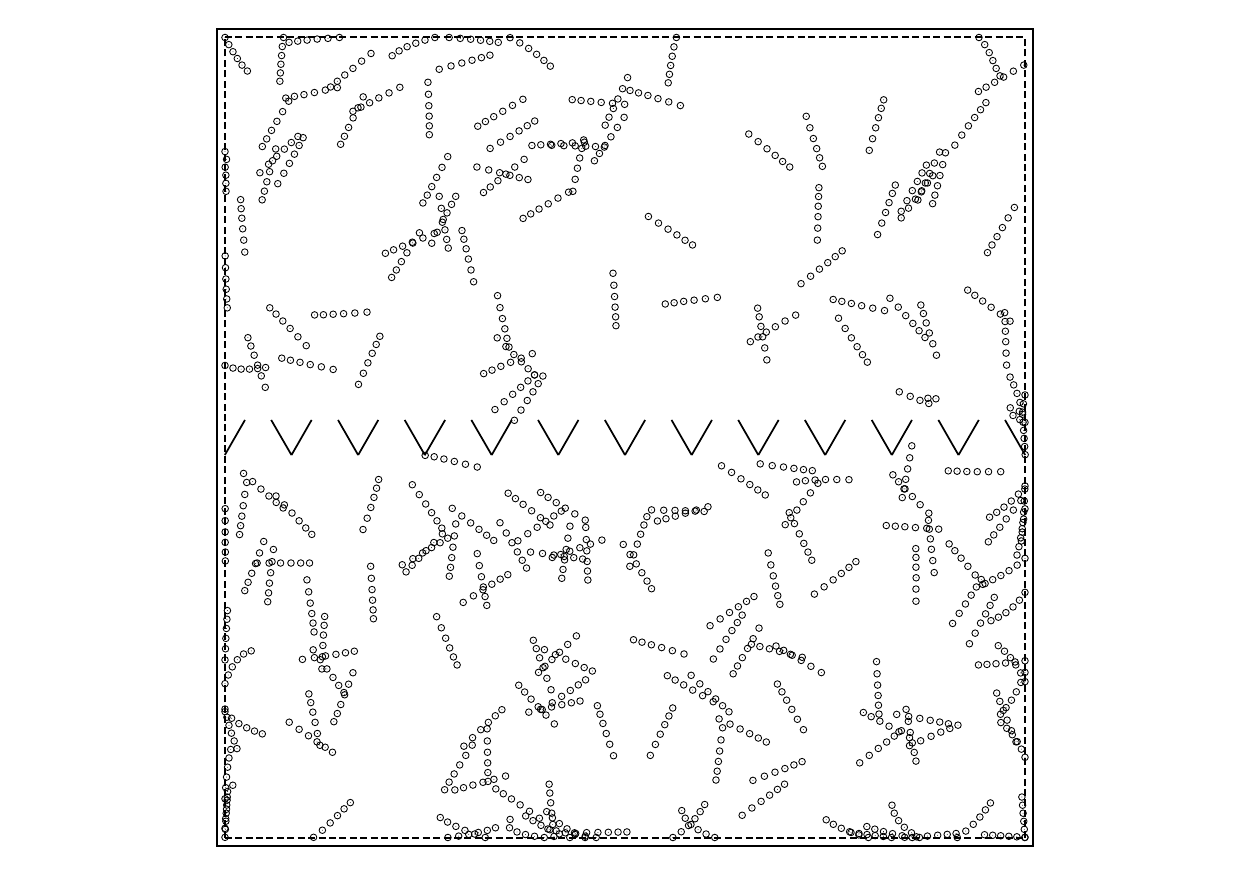}
}
\caption{Simulation box for the six-monomer polymer case undergoing elastic collisions with funnel walls at time, (a): $t=0$, and (b): $t=6\times 10^6$.}
\label{fig:2box6}
\end{figure*}

\begin{figure*}
\begin{center}
\includegraphics[scale=0.3]{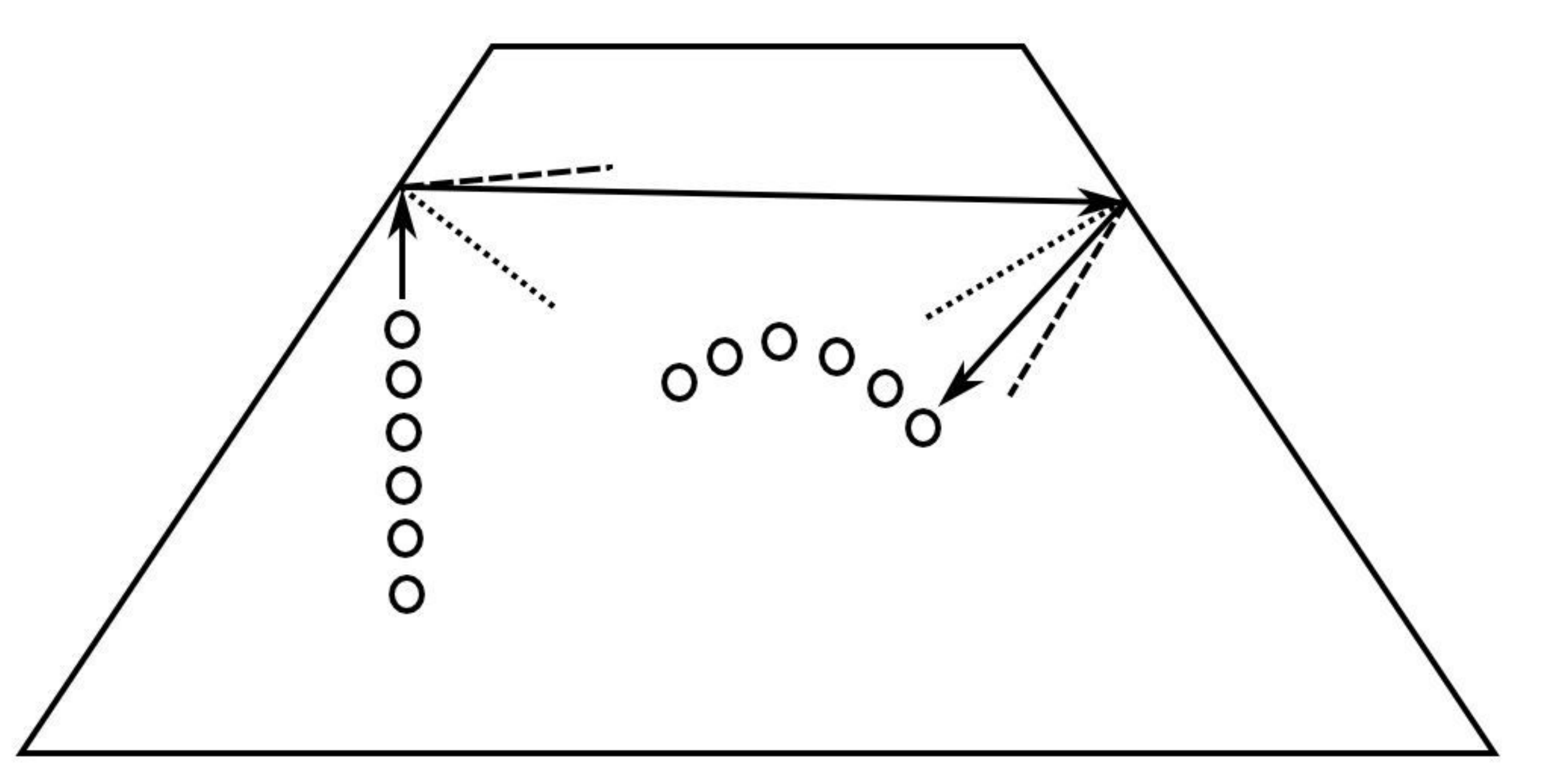}
\caption{Geometry of the space between two funnels depicting the trajectory of a polymer colliding elastically with one of the barriers. The dashed lines represent the trajectories of the first particle of the polymer following Snell's Law, with the dotted lines being the normals to the barriers. The solid lines depict the biased trajectories of the first particle as it is pulled from behind by the spring forces of the following particles.}
\label{fig:bdelas}
\end{center}
\end{figure*}

\begin{figure*}
\begin{center}
\includegraphics[scale=0.85]{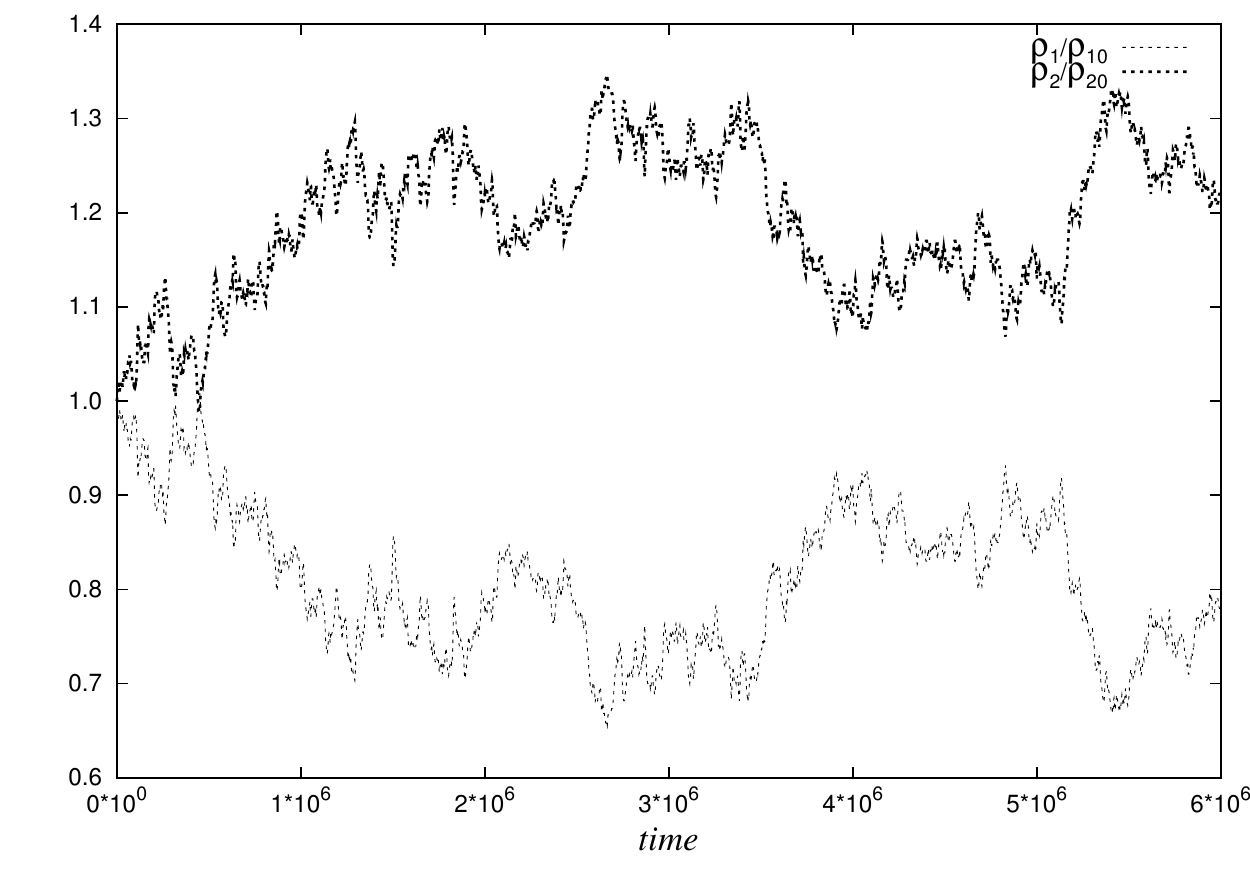}
\caption{Evolution of normalized particle densities, $\rho_1/\rho_{10}$ for chamber $1$ and $\rho_2/\rho_{20}$ for chamber $2$ for the six-monomer polymer system with elastic collisions. The dotted line represents the particle density in chamber 2 whereas the solid line represents that in chamber 1.}
\label{fig:recel} 
\end{center}
\end{figure*}

One significant difference between an elongated chain and an individual particle is that an entropic force is induced when the elongated chain interacts with the anisotropic funnel. As the chains cross the opening, the part of the chain that belongs to chamber 1 has more conformal degrees of freedom than the part in chamber 2. As a result, the chain is pushed to chamber 1. The longer the chain is, the higher the possibility is for the part of the chain in chamber 1 to be longer, giving an increased entropic force. 
This increases the rectification as we increase the number of monomers in the chain or increase the bond length between the monomers.  
The dependence of the conformal entropy on the chain length goes as $(N_{mon}-1)l \sim l^2$.
In the case of increasing the barrier tilting angle, similar to the monomer case, the entropic force of the polymers is enhanced due to the increased probability of the polymers interacting with the barriers.
The entropic force also ensures that the rectification magnitude is enhanced compared to the monomer case, due to more conformal degrees of freedom in the chains.

To demonstrate this entropic contribution as a force that pushes the chains to the upside of the funnels, we remove the ballistic force of the chains. Instead, the bath of polymers is subjected only to a diffusive thermal force described by ${\bf F}_{Tl}$, 
where $\langle F_{Tl} \rangle=0$ and $\langle F_{Tl}(t)F_{Tj}(t') \rangle=2\eta k_B T\delta_{lj}\delta(t-t')$, with $k_B$ being the Boltzmann constant.  
A bath of $200$ polymers with six monomers each are simulated over $6$ million time steps of $dt=0.0005$. A funnel system similar to the cases above is set up with $\theta=45$ degrees. Interestingly, we observe that even without the ballistic force, in contrast with the single-monomer case, a thermal force with $\gamma^2=2\eta k_B T=1$ induces a rectification of $1.2074$ for these six-monomer polymers. Fig.~\ref{fig:rtherm} shows the magnitude of the rectification as $\gamma^2$ is increased from $1$ to $10$. A higher rectification is observed when the thermal force, that is, when the temperature, is increased. With $\theta=45$ degrees, the $l_o$ is set at $1.18$, which is comparable with the total equilibrium length of the chain, $(N_{mon}-1)l=5$ in this case. It should be noted that the rectification disappears if $l_o$ is much larger than the chain length. 

On the other hand, when elastic collisions between the particles with the barrier walls are considered, a \textit{reversed} rectification is observed.  
Fig.~\ref{fig:2box6} shows the simulation box for a system of six-monomer polymers at the initial time and after 6 million time steps. The ballistic force is given to each first particle of every polymer, as $|{\bf F}_m|=2N_{mon}=12$, similar to the system shown in Fig.~\ref{fig:2box}. The run length is also similarly set at $l_r=10$ with a time step of $dt=0.0005$. The barrier setup is the same as that used in the above cases. When the first particle in the polymer hits the barrier wall, it reflects from the wall following Snell's Law. However, its effective post-collision trajectory is ultimately biased such that the angle of reflection is always less than its incident angle (Fig.~\ref{fig:bdelas}). This bias is caused by the pulling spring force of the particle following behind the first particle. As a result, the probability of the polymers to cross from the downside to the upside of the funnels is reduced as compared to their probability to cross from the upside to the downside.  
Fig.~\ref{fig:recel} shows the evolution of the normalized particle densities in chambers 1 and 2 for this six-monomer polymer case. The magnitude of the reversed rectification is $1.653$.
The magnitude of the reversed rectification is reduced when the number of monomers is reduced. This is expected as the magnitude of the pulling spring force decreases proportionately with the number of particles following behind the first particle. When the pulling spring force decreases in magnitude, the bias in the post-collision trajectory of the first particle decreases as well. 

\section{IV. Conclusions}

In this work, we studied the rectification behavior of a bath of polymers performing run-and-tumble dynamics immersed in a two-dimensional box with a system of funnel-shaped barriers.   
Similar to the single-monomer case, the polymers accumulate on the side of the box to which the funnels open up.
As we change the opening angle of the funnels, the rectification behavior of the polymers change in a manner similar to the single-monomer case, that is, the rectification increases as we increase the opening angle.

However, we found that the rectification of the polymers over the asymmetric ratchet system is enhanced compared to the single-monomer case.  
The rectification magnitude increases proportionately with the number of particles composing the polymers.
Correspondingly, the rectification magnitude increases as we increase the bond length between the particles in the polymers.  
We proposed that the additional degree of freedom intrinsic in the non-negligible aspect ratio of the elongated polymers plays a role in the enhancement.   

We confirmed this explanation by performing similar simulations of the systems of polymers but by removing the ballistic forces of the polymers and instead allowing the polymers to undergo only diffusive thermal fluctuations.
With such a system, we observed that in contrast with the single-monomer case which exhibits no rectification, the polymer case shows significant rectification even with small thermal fluctuations.
As we increase the magnitude of the fluctuations, the rectification is seen to increase.
In a separate simulation with ballistic forces, we replaced the elastic collisions with inelastic ones.
In this case, we observed a \textit{reversed} rectification. This is due to the fact that the elongated polymers attempting to cross the barrier system from the downside of the funnels are biased
such that they tend to stay on the downside of the funnels. 
This contrasts with the single-monomer case shown in Refs.~\cite{Cates09,Reichhardt11} where the rectification vanishes when the collisions of the particles with the barriers are elastic.

This modification of the rectification dynamics by the non-negligible aspect ratio of the polymers implies that rectification or work done by baths of soft matter in the presence of asymmetric ratchet systems can be further controlled via the internal structure of the soft matter employed. 
In addition, it also implies that in real systems, the aspect ratio of different species of bacteria may play a role in modifying their rectification magnitudes.  

\section{V. Acknowledgements} 

YSJ acknowledges the Max Planck
Society (MPG), the Korea Ministry of Education, Science
and Technology (MEST), Gyeongsangbuk-Do and Pohang
City for the support of the Independent Junior Research Group
at the Asia-Pacific Center for Theoretical Physics (APCTP),
as well as the National Research Foundation of
Korea under the grant funded by the Korean government (MEST) (NRFC1ABA001-
2011-0029960, and 2012R1A1A2009275).

\bibliographystyle{prsty}
\bibliography{references}

\end{document}